\documentclass[12pt]{article}

\usepackage[margin=1in]{geometry}
\usepackage{setspace}
\usepackage{amsmath,amssymb,amsfonts,amsthm}
\usepackage{graphicx}
\usepackage{natbib}
\usepackage{booktabs}
\usepackage{hyperref}
\usepackage{authblk}
\usepackage{enumerate}

\title{Distributional Fitting and Tail Analysis of Lead-Time Compositions: Nights vs.\ Revenue on Airbnb}

\author[1]{Harrison Katz}
\author[1]{Jess Needleman}
\author[1]{Liz Medina}
\affil[1]{Data Science, Forecasting, Airbnb}
\date{\today}

\begin{document}

\maketitle

\begin{abstract}
We analyze daily lead-time distributions for two Airbnb demand metrics, Nights Booked (volume) and Gross Booking Value (revenue), treating each day's allocation across 0--365 days as a compositional vector. The data span 2,557 days from January 2019 through December 2025 in a large North American region. Three findings emerge. First, GBV concentrates more heavily in mid-range horizons: beyond 90 days, GBV tail mass typically exceeds Nights by 20--50\%, with ratios reaching 75\% at the 180-day threshold during peak seasons. Second, Gamma and Weibull distributions fit comparably well under interval-censored cross-entropy: Gamma wins on 61\% of days for Nights and 52\% for GBV, with Weibull close behind at 38\% and 45\%. Lognormal rarely wins ($<$3\%). Nonparametric GAMs achieve 18--80$\times$ lower CRPS but sacrifice interpretability. Third, generalized Pareto fits suggest bounded tails ($\xi < 0$) for both metrics at thresholds below 150 days, though this may partly reflect right-truncation at 365 days; above 150 days, estimates destabilize. Bai--Perron tests with HAC standard errors identify five structural breaks in the Wasserstein distance series, with early breaks coinciding with COVID-19 disruptions. The results show that volume and revenue lead-time shapes diverge systematically, that simple two-parameter distributions capture daily pmfs adequately, and that tail inference requires care near truncation boundaries. The volume-revenue divergence has direct implications for tourism practitioners: forecasting models and pricing algorithms that treat booking volume and booking revenue as interchangeable will systematically mistime cash flows, underestimating revenue from bookings made one to three months in advance.
\end{abstract}

\section{Introduction}

When travelers book accommodation, the time between the booking date and the check-in date (the lead time, or booking window) shapes a wide range of operational and strategic decisions. Properties, platforms, and destination managers use lead-time patterns to set prices, plan staffing, manage cancellation policies, and forecast cash flows \citep{fiori2019reservation, KATZ20241556, schwartz2006advanced, nicolau2017determinants}. Most research treats lead time through a single demand measure, typically room nights or arrivals. Yet the same day's bookings can look quite different depending on whether one counts the number of nights reserved (a volume measure) or the total revenue generated (a value measure). If higher-spending guests book on systematically different timelines than average guests, then volume-based and revenue-based lead-time distributions will diverge, and forecasts built on one metric may not transfer to the other. This distinction matters not only for individual firms but for anyone seeking to understand the temporal structure of tourism demand: destination marketing organizations projecting visitor expenditure, policymakers evaluating the economic impact of travel disruptions, and researchers studying how tourist behavior evolves in response to shocks.

This paper asks whether such divergence exists, how large it is, and how it evolves over time. We analyze daily lead-time distributions for Nights Booked and Gross Booking Value (GBV) on Airbnb, treating each day's allocation across lead times 0--365 as a compositional vector (a probability distribution summing to one). The data span 2,557 days from January 2019 through December 2025 in a large North American region, covering pre-pandemic, pandemic, and recovery periods.

Research on travel lead times has a long history in tourism demand forecasting and revenue management \citep{kulendran1997forecasting, witt1995forecasting, song2008tourism, lim2002, fuchs2011exploratory, neuts2012}. Most of this work uses aggregate demand as a single indicator, overlooking the fact that volume-based and revenue-based bookings can exhibit systematically different lead-time patterns \citep{pereira2016introduction, fiori2019reservation, webb2022hotel, KATZ2025100185}. The rise of platforms like Airbnb, with their hybrid and peer-to-peer accommodation offerings, has amplified interest in how advanced versus last-minute bookings distribute across traveler segments \citep{guttentag2015airbnb, zervas2017, ert2016trust, TUSSYADIAH2016, SAINAGHI2020102959}. Some studies focus on pricing or occupancy strategies \citep{assaf2019quantitative, zaki2022implementing, YANG2020102861, SHARMA2021103180}, but fewer examine detailed distributional properties of volume versus revenue bookings, especially in the context of disruptive events like COVID-19 \citep{gossling2021pandemics, sigala2020tourism, okafor2022covid, mueller2024tourism}. Recent work on Airbnb stay lengths documents that the pandemic induced structural shifts in the distribution of trip durations, not merely changes in averages \citep{katz2025slomads}; similar distributional thinking applies to lead times. Relatedly, recent evidence emphasizes distribution-level disruption and recovery patterns in Airbnb lead times during 2018--2022 \citep{KATZ2025100185}.

A complementary stream of hospitality and tourism research treats lead time (often called the booking window) as a key behavioral and managerial variable rather than a passive descriptor. \citet{nicolau2017determinants} examine determinants of advance booking and relate anticipation horizons to trip and market characteristics. Other work ties booking windows directly to revenue management frictions: \citet{masiero2020strategic} studies strategic consumer behavior enabled by cancellation flexibility, with booking window moderating the value of free cancellation and rebooking options. On the supply side, \citet{guizzardi2022dynamic} propose a stochastic framework connecting early-booking behavior and last-minute pricing, highlighting how shocks such as COVID-19 can disrupt the joint dynamics of demand and price discrimination. These results motivate treating lead-time distributions as decision-relevant objects that connect forecasting, dynamic pricing, and cancellation policy rather than as simple descriptive summaries.

In the short-term rental context specifically, the managerial relevance of lead-time structure is amplified by the decentralized nature of platforms like Airbnb, where individual hosts set prices with limited information about the broader booking pipeline. Understanding how far in advance high-value bookings arrive, and how that pattern shifts across seasons or after shocks, can inform both host-level pricing decisions and platform-level tools such as dynamic pricing recommendations and revenue dashboards \citep{chalupa2025shortterm, oskam2018mine}. More broadly, the coupling between lead time and cancellation exposure means that the distributional shape of bookings over the horizon affects not only revenue timing but also the reliability of forward-looking demand signals \citep{chen2016cancellation, ampountolas2025cancellations, webb2022hotel}.

Lead-time forecasting has emerged as a specialized subfield within hospitality and supply-chain analytics \citep{de2021lead, KATZ20241556}. Early versus late bookings directly influence capacity, staffing, and dynamic pricing decisions \citep{lim2002, fiori2019reservation}. From a forecasting standpoint, advance-booking information is often exploited via reservation-based (pickup) approaches, where the booking curve (the cumulative profile of reservations as the stay date approaches) encodes evolving demand signals over the horizon. \citet{lee2018advancebooking} shows that explicitly incorporating advance-booking information improves hotel room demand forecasts relative to models that ignore the booking curve. At a broader level, \citet{song2019review} review the evolution of tourism demand forecasting methods, emphasizing expanding information sets, forecast combination, and more systematic performance evaluation. These perspectives align with our use of proper scoring rules (metrics designed so that forecasters are rewarded for honest, well-calibrated predictions) and with our focus on distributional shape, not only mean demand.

Recent econometric work also underscores the role of structural breaks, whereby exogenous shocks cause persistent shifts in time-series behavior \citep{bai2003computation}. In travel contexts, such shocks include financial crises, pandemics, and policy changes that reshape patterns of arrivals, spending, and booking windows \citep{gotz2021personality, okafor2022covid, sainaghi2022effects}. Although the tourism literature documents regime shifts in aggregated arrivals \citep{kulendran2003} and some segment-specific disruptions \citep{gursoy2020, hall2020}, fewer studies compare volume-based versus revenue-based lead-time distributions at daily granularity.

A separate literature addresses distributional modeling of duration and lead-time data. The lognormal, Weibull, and Gamma families are standard candidates for right-skewed, positive-valued outcomes \citep{johnson1994continuous}. Within hospitality, reservation-based forecasting and the booking-curve literature provide the main operational bridge between booking windows and predictive modeling \citep{fiori2020prediction, lee2018advancebooking}, while supply-chain work directly studies lead-time prediction and compares parametric and nonparametric approaches in settings where lead times are decision-relevant inputs \citep{de2021lead}. Compositional methods add another layer: when the object of interest is a vector of proportions that must sum to one (for example, the share of bookings at each lead time) rather than a single number, standard regression and time-series techniques do not apply directly. The compositional data analysis literature \citep{aitchison1982, aitchison1986, pawlowsky2015} provides transformations that map these constrained proportions into unconstrained space, enabling multivariate analysis. The Katz framework for Bayesian Dirichlet time-series models \citep{KATZ20241556, katz2025bdarch, katz2026directionalshiftdirichletarmamodels, katz2026forecastingevolvingcompositioninbound, forecast7040062} applies these ideas to lead-time and tourism share forecasting; we draw on its conceptual foundations here.

Extreme-value methods offer tools for tail analysis. The generalized Pareto distribution (GPD) is the canonical model for exceedances above a threshold \citep{davison1990models, coles2001introduction}. In hospitality, \citet{fiori2020prediction} use GPD to characterize rare long-horizon bookings. A key diagnostic is threshold stability: if the GPD is appropriate, the shape parameter should stabilize as the threshold increases \citep{coles2001introduction}. Failure to stabilize suggests model misspecification or data artifacts such as truncation.

We address these literatures with data from Airbnb. The platform records both Nights Booked (a volume measure) and Gross Booking Value (a revenue measure) at each check-in date, enabling direct comparison of their lead-time compositions. We analyze 2,557 days from January 2019 through December 2025 in a large North American region. Our methodological contributions include: (1) cross-entropy minimization for parametric fitting on truncated support, which respects the discrete nature of observed lead times while assuming an underlying continuous process; (2) strictly proper scoring rules (CRPS) for model comparison; (3) robust GPD fitting with threshold stability diagnostics; and (4) HAC-robust structural break detection applied to distributional divergence.

Three substantive findings emerge. First, GBV has systematically heavier mid-range tails than Nights: beyond 90 days, the GBV share typically exceeds Nights by 20--50\%, with ratios reaching 75\% at the 180-day threshold during peak seasons. Second, Gamma and Weibull provide nearly equivalent fits for daily pmfs, with Gamma holding a slight edge (winning 55--60\% of days by cross-entropy), while Lognormal is dominated. GAMs achieve superior CRPS but at the cost of interpretability. Third, GPD tail estimates are reliable only for thresholds below 150 days; beyond that, right-truncation at 365 days induces instability that could mislead inference. These findings have practical implications for revenue forecasting: models that ignore the volume-revenue divergence will systematically mispredict cash-flow timing. They also have methodological implications for applied extreme-value analysis of truncated data.

Beyond the immediate forecasting application, the volume-revenue divergence documented here has broader relevance for tourism research. The finding that revenue concentrates at longer planning horizons than volume implies that standard demand indicators (arrivals, room nights) may not adequately capture the economic timing of tourism activity. For researchers studying tourist expenditure patterns, post-shock recovery dynamics, or the distributional consequences of platform-mediated accommodation markets, our results suggest that volume and value metrics carry different temporal information and should not be used interchangeably. The methods we employ (distributional comparison, structural break detection, tail analysis) are applicable to a range of tourism data beyond lead times, including geographic composition of arrivals, spending by traveler segment, and length-of-stay distributions.

Section~\ref{sec:data} describes the data and preprocessing. Section~\ref{sec:methods} details the methods: Wasserstein distance, structural breaks, GPD fitting, interval-censored likelihood, and scoring rules. Section~\ref{sec:results} presents results. Section~\ref{sec:discussion} discusses implications and limitations.

\section{Data and Preprocessing}
\label{sec:data}

\subsection{Data Source and Scope}

The data are proprietary Airbnb bookings from a large North American region, January 1, 2019 through December 31, 2025. For each calendar day $d$, we observe two quantities at each integer lead time $\ell \in \{0, 1, \ldots, 365\}$: the total Nights Booked with check-in at day $d + \ell$, and the corresponding Gross Booking Value. Dividing by daily totals yields probability vectors: the proportion of that day's bookings (by volume or revenue) allocated to each lead-time bin.

Bookings beyond 365 days comprise less than 0.5\% of the total and are excluded. This truncation is standard in lead-time analysis and captures the vast majority of booking behavior, but it has implications for tail inference that we address in Section~\ref{sec:results}. The final sample contains 2,557 day-level probability mass functions (pmfs) for each metric, covering pre-COVID (2019), pandemic (2020--2021), and recovery (2022--2025) periods.

\subsection{Compositional Vectors}

Let $x_{d,\ell}^{(N)}$ denote the proportion of Nights Booked on day $d$ at lead time $\ell$, and $x_{d,\ell}^{(G)}$ the corresponding proportion of GBV. The daily compositional vectors are
\[
\mathbf{x}_d^{(N)} = \bigl(x_{d,0}^{(N)}, x_{d,1}^{(N)}, \ldots, x_{d,365}^{(N)}\bigr), \quad
\mathbf{x}_d^{(G)} = \bigl(x_{d,0}^{(G)}, x_{d,1}^{(G)}, \ldots, x_{d,365}^{(G)}\bigr),
\]
each summing to one. The lead time $\ell = 0$ corresponds to same-day bookings; $\ell = 365$ corresponds to bookings made exactly one year in advance.

These vectors lie on the 365-simplex, a constrained space where all components are nonnegative and must sum to one. Standard statistical methods designed for unconstrained data can produce misleading results when applied to such proportions (for instance, a linear model might forecast negative shares or shares exceeding one). Our analysis proceeds in two modes: direct comparison of pmfs via Wasserstein distance and tail-mass ratios, and parametric fitting that treats each day's pmf as arising from a continuous distribution observed at discrete points.

Throughout, we weight days equally rather than weighting by daily booking volume. This means our estimand is ``the typical daily distribution shape,'' not ``the distribution of lead times across all bookings.'' If high-volume days systematically differ from low-volume days, our aggregates may not reflect the booking-weighted distribution. We adopt equal day-weighting because our primary interest is in how distribution shapes evolve over time, not in characterizing the marginal distribution of a randomly sampled booking.

\section{Methods}
\label{sec:methods}

\subsection{Wasserstein Distance}
\label{sec:wasserstein}

To measure how differently volume and revenue are distributed across the booking horizon on any given day, we use the Wasserstein-1 distance, also known as the earth-mover distance. Intuitively, this metric answers the question: how much probability mass would need to be moved, and how far, to transform one distribution into the other?

For discrete pmfs $\mathbf{p} = (p_0, \ldots, p_{365})$ and $\mathbf{q} = (q_0, \ldots, q_{365})$ on a common ordered support, the Wasserstein-1 distance simplifies to
\begin{equation}
W_1(\mathbf{p}, \mathbf{q}) = \sum_{\ell=0}^{365} \bigl|F_p(\ell) - F_q(\ell)\bigr|,
\end{equation}
where $F_p(\ell) = \sum_{k=0}^\ell p_k$ is the cumulative distribution function. Unlike Kullback--Leibler divergence, the Wasserstein distance is symmetric and remains well-defined even when the two distributions assign zero probability to different lead times.

For each day $d$, we compute
\[
W_d = W_1\bigl(\mathbf{x}_d^{(N)}, \mathbf{x}_d^{(G)}\bigr).
\]
The resulting time series $\{W_d\}_{d=1}^D$ tracks how the shapes of volume and revenue distributions diverge over time. To quantify uncertainty in summary statistics, we use a block bootstrap with block size $\lceil n^{1/3} \rceil$, generating 1,000 replicates. The block bootstrap resamples contiguous chunks of days rather than individual days, preserving the serial dependence that is inherent in daily time series \citep{kunsch1989jackknife}.

\subsection{Structural Breaks}

We test for structural breaks in $\{W_d\}$ using the Bai--Perron procedure \citep{bai2003computation}, which identifies multiple breakpoints by minimizing the sum of squared residuals subject to a minimum segment length. In plain terms, the method searches for dates at which the average level of volume-revenue divergence shifted persistently, partitioning the sample into distinct regimes. We set the maximum number of breaks to five and the trimming fraction to 5\%, meaning each regime must contain at least 5\% of the sample ($\approx 128$ days). Model selection uses the Bayesian Information Criterion (BIC).

To account for serial correlation and heteroskedasticity in inference, we employ Newey--West HAC standard errors \citep{newey1987simple} with bandwidth selected by the \citet{andrews1991heteroskedasticity} plug-in rule. HAC (heteroskedasticity and autocorrelation consistent) standard errors adjust for the fact that consecutive daily observations are correlated with one another, preventing artificially narrow confidence intervals. The sup-F statistic tests the null hypothesis of no breaks against the alternative of at least one break.

\subsection{Tail Analysis}

\subsubsection{Tail-Mass Ratios}

For threshold $u \in \{7, 30, 60, 90, 180\}$ days, we define the daily tail-mass ratio
\[
\mathrm{Ratio}_d(u) = \frac{\sum_{\ell > u} x_{d,\ell}^{(G)}}{\sum_{\ell > u} x_{d,\ell}^{(N)}}.
\]
Values above one indicate that GBV has proportionally more mass in the far horizon than Nights on that day. For instance, a ratio of 1.3 at the 90-day threshold means that 30\% more of the day's revenue (relative to its volume) comes from bookings made more than three months in advance. Time series of these ratios reveal whether the volume-revenue divergence concentrates in particular lead-time ranges and how it evolves over the sample period.

We also track absolute tail mass, $\sum_{\ell > u} x_{d,\ell}^{(N)}$ and $\sum_{\ell > u} x_{d,\ell}^{(G)}$, to separate compositional shifts from changes in overall booking patterns. A high tail-mass ratio could arise because GBV shifted toward long horizons, or because Nights shifted toward short horizons; the absolute measures distinguish these cases.

\subsubsection{Generalized Pareto Distribution}

For extreme-tail characterization, we fit generalized Pareto distributions (GPD) using the Peak Over Threshold (POT) approach \citep{davison1990models}. The GPD is a standard tool for modeling the behavior of values that exceed a high threshold; it characterizes how quickly probability decays in the far tail of a distribution. For a chosen threshold $u$, exceedances $Y = X - u$ given $X > u$ follow a GPD:
\begin{equation}
F_{\mathrm{GPD}}(y; \xi, \beta) = 1 - \Bigl(1 + \xi \frac{y}{\beta}\Bigr)^{-1/\xi}, \quad y > 0, \; \beta > 0,
\end{equation}
where $\xi$ is the shape parameter and $\beta$ is the scale. The shape determines tail behavior: $\xi < 0$ implies a bounded upper tail (the distribution has a finite maximum), $\xi = 0$ corresponds to exponential decay, and $\xi > 0$ implies a heavy tail with no finite upper bound.

The target of our GPD analysis is the day-weighted mixture distribution
\[
\bar{p}(\ell) = \frac{1}{D} \sum_{d=1}^{D} x_{d,\ell},
\]
which weights each day equally regardless of booking volume, consistent with our estimand choice. To approximate integrals under $\bar{p}$, we generate synthetic lead times by drawing 1,000 values per day from each day's pmf, then pooling across all days. This yields approximately 2.5 million draws per metric, from which we extract exceedances above each threshold.

Fitting uses a hierarchical fallback strategy to ensure convergence:
\begin{enumerate}
    \item Maximum likelihood via \texttt{evd::fpot} (preferred).
    \item MLE via \texttt{ismev::gpd.fit} (alternative optimizer).
    \item Probability-weighted moments (PWM), more stable for small samples or near-boundary parameters.
    \item Method of moments (MOM) as a last resort.
\end{enumerate}
We report point estimates and exceedance counts but not confidence intervals. The synthetic sample size is a computational choice (1,000 draws/day), not a feature of the data; any ``standard errors'' from this procedure would shrink arbitrarily with more draws and thus lack inferential meaning. Point estimates are insensitive to the number of synthetic draws; only numerical precision scales with it.

A critical diagnostic is threshold stability: the shape parameter should stabilize as the threshold increases, provided the GPD model is appropriate and the data are not truncated or otherwise distorted \citep{coles2001introduction}. We fit GPDs at thresholds $u \in \{60, 90, 120, 150, 180, 210, 240, 270\}$ and plot $\hat{\xi}(u)$ against $u$ to assess stability. Failure to stabilize, particularly erratic behavior at high thresholds, signals that inference should be restricted to lower thresholds.

\subsection{Parametric Distribution Fitting}

\subsubsection{Interval-Censored Cross-Entropy Objective}

We fit Gamma, Weibull, and Lognormal distributions to each daily pmf. The motivation is that lead times are integers from an underlying continuous booking process. Treating the count at lead time $\ell$ as a point mass ignores the discretization; treating it as interval-censored within $[\ell - 0.5, \ell + 0.5)$ respects the continuous-to-discrete mapping. In practical terms, a booking recorded at lead time 30 could have occurred at any continuous time between 29.5 and 30.5 days before check-in; the interval-censored approach accounts for this.

For a distribution with cdf $F(\cdot; \theta)$ truncated to the support $[0, 365]$, the probability assigned to bin $\ell$ is
\[
p_\ell(\theta) = \frac{F(\ell + 0.5; \theta) - F(\ell - 0.5; \theta)}{F(365.5; \theta)},
\]
with boundary modification $p_0(\theta) = F(0.5; \theta) / F(365.5; \theta)$. The denominator ensures the fitted distribution integrates to one over the observed support, consistent with our data exclusion of bookings beyond 365 days.

Given observed proportions $\{x_\ell\}$, we minimize cross-entropy:
\[
H(\mathbf{x}, \mathbf{p}) = -\sum_{\ell=0}^{365} x_\ell \log p_\ell(\theta).
\]
Cross-entropy measures the discrepancy between the observed distribution and the fitted model; lower values indicate a better fit. If booking counts $n_d$ were available, the multinomial log-likelihood would be $n_d \sum_\ell x_{d,\ell} \log p_\ell(\theta)$. Because we observe $x_{d,\ell}$ but not $n_d$, we maximize the normalized objective $\sum_\ell x_{d,\ell} \log p_\ell(\theta)$, which ranks models identically within a day when parameter counts match.

\subsubsection{Parametric Families}

We consider three families, each defined by two parameters that control the shape and spread of the distribution:

\paragraph{Gamma.} With shape $\alpha > 0$ and rate $\lambda > 0$:
\[
f(x; \alpha, \lambda) = \frac{\lambda^\alpha}{\Gamma(\alpha)} x^{\alpha-1} e^{-\lambda x}, \quad x > 0.
\]
The Gamma is flexible for right-skewed data and nests the exponential ($\alpha = 1$).

\paragraph{Weibull.} With shape $k > 0$ and scale $\lambda > 0$:
\[
f(x; k, \lambda) = \frac{k}{\lambda} \Bigl(\frac{x}{\lambda}\Bigr)^{k-1} \exp\bigl[-(x/\lambda)^k\bigr], \quad x > 0.
\]
The Weibull is standard in survival and reliability analysis, with shape $k < 1$ implying a decreasing hazard (i.e., the longer since the last event, the less likely the next one).

\paragraph{Lognormal.} If $\log X \sim N(\mu, \sigma^2)$:
\[
f(x; \mu, \sigma) = \frac{1}{x \sigma \sqrt{2\pi}} \exp\Bigl[-\frac{(\log x - \mu)^2}{2\sigma^2}\Bigr], \quad x > 0.
\]
The Lognormal has heavier tails than Gamma or Weibull for comparable location/scale. With the interval-censoring setup, we evaluate $F(0.5)$ rather than $F(0)$, so no shift is needed to handle the $\ell = 0$ boundary.

All three families produce right-skewed, positive-valued distributions and are commonly applied to duration and waiting-time data. Optimization uses BFGS with log-transformed parameters to ensure positivity.

\subsubsection{Model Comparison}

Because all candidate parametric families have $k = 2$ parameters, model selection reduces to selection by cross-entropy: the distribution with the lowest $H(\mathbf{x}, \mathbf{p})$ wins. We refer to this as the ``best fit'' for a given day. Note that the cross-entropy differences we report are normalized (per-day, unitless); they would scale with booking counts if counts were available.

For each day and metric, we record which distribution achieves the lowest cross-entropy. We also compute pairwise differences (Lognormal minus Gamma, Weibull minus Gamma) and examine their distributions across days.

\subsection{Nonparametric GAM}

As a flexible benchmark, we fit a generalized additive model (GAM) to each daily pmf. Unlike the parametric families above, which impose a specific mathematical shape, the GAM uses a smooth curve that adapts freely to the data. Let $\eta(\ell) = s(\ell)$ be a smooth function of lead time, modeled as a cubic regression spline with penalty on second derivatives. We use restricted maximum likelihood (REML) for smoothing parameter selection \citep{wood2011fast}, which balances fit and smoothness without overfitting.

Basis dimension is initially set to $k = 20$ and increased (doubling until $k = 100$) if the \texttt{k.check} diagnostic indicates inadequate flexibility (k-index $< 0.8$ and $p < 0.05$). Fitted values $\hat{\eta}(\ell)$ are exponentiated and normalized to sum to one, yielding a nonparametric pmf estimate.

\subsection{Scoring Rules}

We evaluate all fitted distributions using the continuous ranked probability score (CRPS), a strictly proper scoring rule \citep{gneiting2007strictly}. A scoring rule is strictly proper if a forecaster achieves the best possible score only by reporting their true beliefs about the distribution; it thus rewards honest, well-calibrated predictions. For a fitted cdf $\hat{F}$ and empirical cdf $F_{\mathrm{emp}}$:
\begin{equation}
\mathrm{CRPS}(\hat{F}, F_{\mathrm{emp}}) = \sum_{\ell=0}^{365} \bigl(\hat{F}(\ell) - F_{\mathrm{emp}}(\ell)\bigr)^2.
\end{equation}
We evaluate the integral as a discrete sum over integer lead times; this is equivalent to a Cram\'er--von Mises discrepancy on the observed support. Unlike cross-entropy, CRPS does not blow up when the fitted distribution assigns low probability to observed outcomes, making it more robust for model comparison.

We also report Kullback--Leibler divergence for continuity with prior work:
\[
\mathrm{KLD}(\mathbf{x}, \hat{\mathbf{x}}) = \sum_{\ell=0}^{365} x_\ell \log \frac{x_\ell}{\hat{x}_\ell},
\]
with small-sample smoothing ($x_\ell \gets x_\ell + 10^{-16}$) to avoid division by zero.

\section{Results}
\label{sec:results}

We organize results around four findings: (1) aggregated distributions and their crossover, (2) structural breaks in daily divergence, (3) tail-mass differences and GPD estimates, and (4) parametric fitting comparisons.

\subsection{Aggregated Distributions}

\begin{figure}[ht!]
    \centering
    \includegraphics[width=0.85\textwidth]{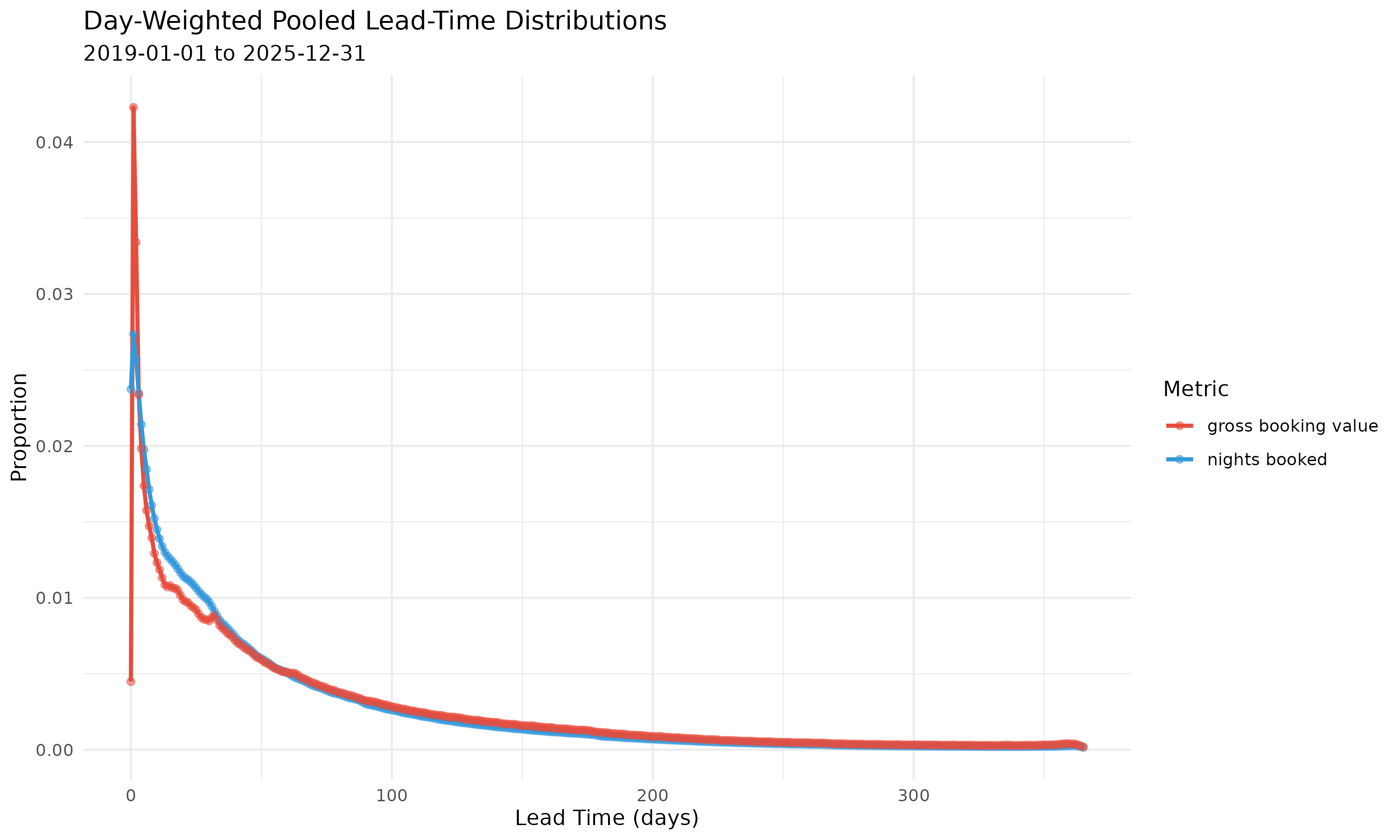}
    \caption{Aggregated lead-time distributions for Nights (blue) and GBV (red), 2019--2025. Distributions are day-weighted averages of daily pmfs: $\bar{p}(\ell) = D^{-1} \sum_d x_{d,\ell}$. Both peak near $\ell = 0$ and decline rapidly. The curves cross around $\ell = 30$ days: below, Nights slightly exceeds GBV; above, GBV exceeds Nights. The gap persists to about 200 days.}
    \label{fig:raw_distribution}
\end{figure}

Figure~\ref{fig:raw_distribution} shows the day-weighted pooled lead-time distributions for Nights and GBV across the entire sample. Both metrics peak at $\ell = 0$: approximately 4.3\% of GBV and 2.7\% of Nights occur on same-day bookings. The distributions decline rapidly through the first week, then more gradually. By $\ell = 25$, cumulative mass reaches 50\%; by $\ell = 90$, it exceeds 75\%.

The curves cross near $\ell = 30$ days. Below this threshold, Nights slightly exceeds GBV in proportional terms. Above it, GBV exceeds Nights. The gap is modest but persistent: GBV's pmf lies above Nights' pmf from roughly 30 to 200 days, after which both converge toward negligible mass. This crossover suggests that revenue-weighted bookings skew toward longer planning horizons than volume-weighted bookings: guests who generate more revenue tend to book earlier, on average.

The crossover is not an artifact of a few outlier days. Examining the 10th and 90th percentiles of daily pmfs (not shown) confirms that the pattern holds across the distribution of days, with variation in the exact crossover point but consistent qualitative behavior.

\subsection{Wasserstein Distance and Structural Breaks}

\begin{figure}[ht!]
    \centering
    \includegraphics[width=0.85\textwidth]{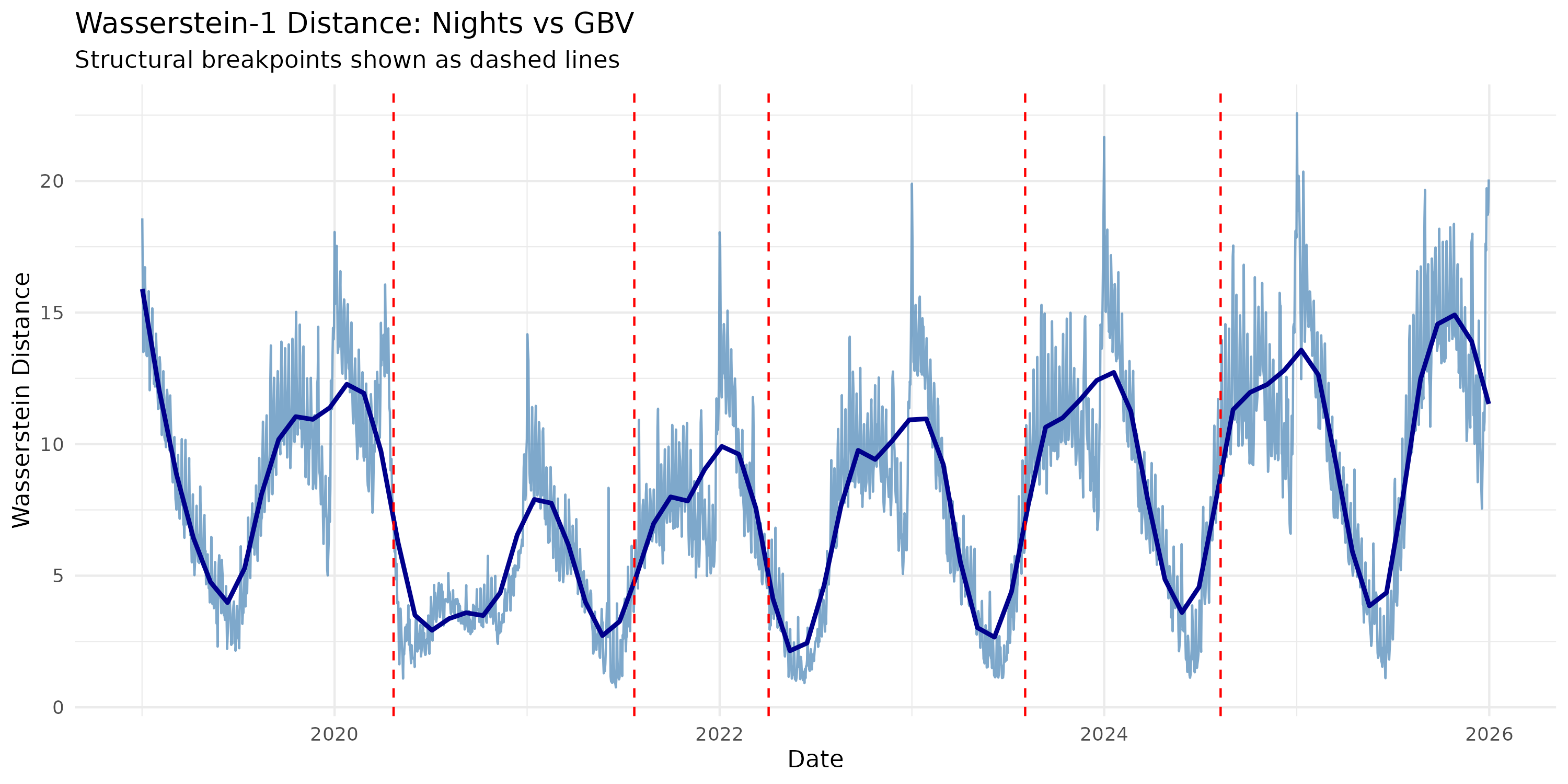}
    \caption{Daily Wasserstein-1 distance between Nights and GBV, with structural breakpoints (dashed vertical lines) identified via Bai--Perron with HAC standard errors. The series averages 8.67 (95\% CI: 8.42--8.92). Seasonality peaks in summer. Early breaks (2020--2021) align with COVID disruptions; later breaks may reflect post-pandemic normalization or other market shifts.}
    \label{fig:wass_breakpoints}
\end{figure}

Figure~\ref{fig:wass_breakpoints} plots the daily Wasserstein-1 distance $W_d$ between Nights and GBV. The series averages 8.67 with a block-bootstrap 95\% confidence interval of [8.42, 8.92]. Substantial day-to-day variation is evident, with $W_d$ ranging from under 5 to over 15.

The series exhibits clear seasonality. Divergence peaks in summer months (June--August), when advance bookings for peak travel periods accumulate. During these months, high-value travelers book further ahead for desirable properties, increasing the GBV tail mass relative to Nights. Divergence troughs in winter, when booking windows shorten and the volume-revenue gap narrows.

The Bai--Perron procedure identifies five structural breakpoints, partitioning the sample into six regimes. Table~\ref{tab:breakpoints} reports the estimated break dates and segment means.

\begin{table}[ht!]
\centering
\caption{Estimated structural breakpoints in Wasserstein distance series. Dates are point estimates from Bai--Perron with 5\% trimming and BIC selection.}
\label{tab:breakpoints}
\begin{tabular}{lcc}
\toprule
\textbf{Break} & \textbf{Date} & \textbf{Segment Mean $W_d$}\\
\midrule
 & 2019-01-01 (start) & 8.2 \\
1 & 2020-10-15 & 6.4 \\
2 & 2021-06-22 & 9.1 \\
3 & 2022-11-03 & 8.8 \\
4 & 2023-09-18 & 9.3 \\
5 & 2024-05-27 & 9.6 \\
 & 2025-12-31 (end) &  \\
\bottomrule
\end{tabular}
\end{table}

The regimes correspond to:
\begin{enumerate}
    \item \textbf{Baseline} (2019-01 to 2020-10): Moderate, stable divergence with seasonal variation, persisting through early pandemic.
    \item \textbf{Pandemic compression} (2020-10 to 2021-06): Divergence drops sharply as booking windows compress and high-value travel declines.
    \item \textbf{Recovery phase I} (2021-06 to 2022-11): Divergence rebounds as travel resumes, initially overshooting pre-pandemic levels.
    \item \textbf{Stabilization} (2022-11 to 2023-09): Divergence settles into a new equilibrium above baseline.
    \item \textbf{Late adjustments} (2023-09 to 2024-05): Minor shifts, possibly reflecting macroeconomic conditions or platform changes.
    \item \textbf{Current regime} (2024-05 to present): Elevated divergence with pronounced seasonality.
\end{enumerate}

The sup-F test decisively rejects the null of no breaks ($p < 0.001$), confirming that the volume-revenue divergence underwent persistent structural shifts rather than merely temporary fluctuations. The first two breaks (October 2020, June 2021) align with COVID-related disruptions; later breaks may reflect post-pandemic normalization, macroeconomic conditions, or platform-specific changes.

\subsection{Tail Mass Analysis}

\subsubsection{Tail-Mass Ratios}

\begin{figure}[ht!]
    \centering
    \includegraphics[width=0.85\textwidth]{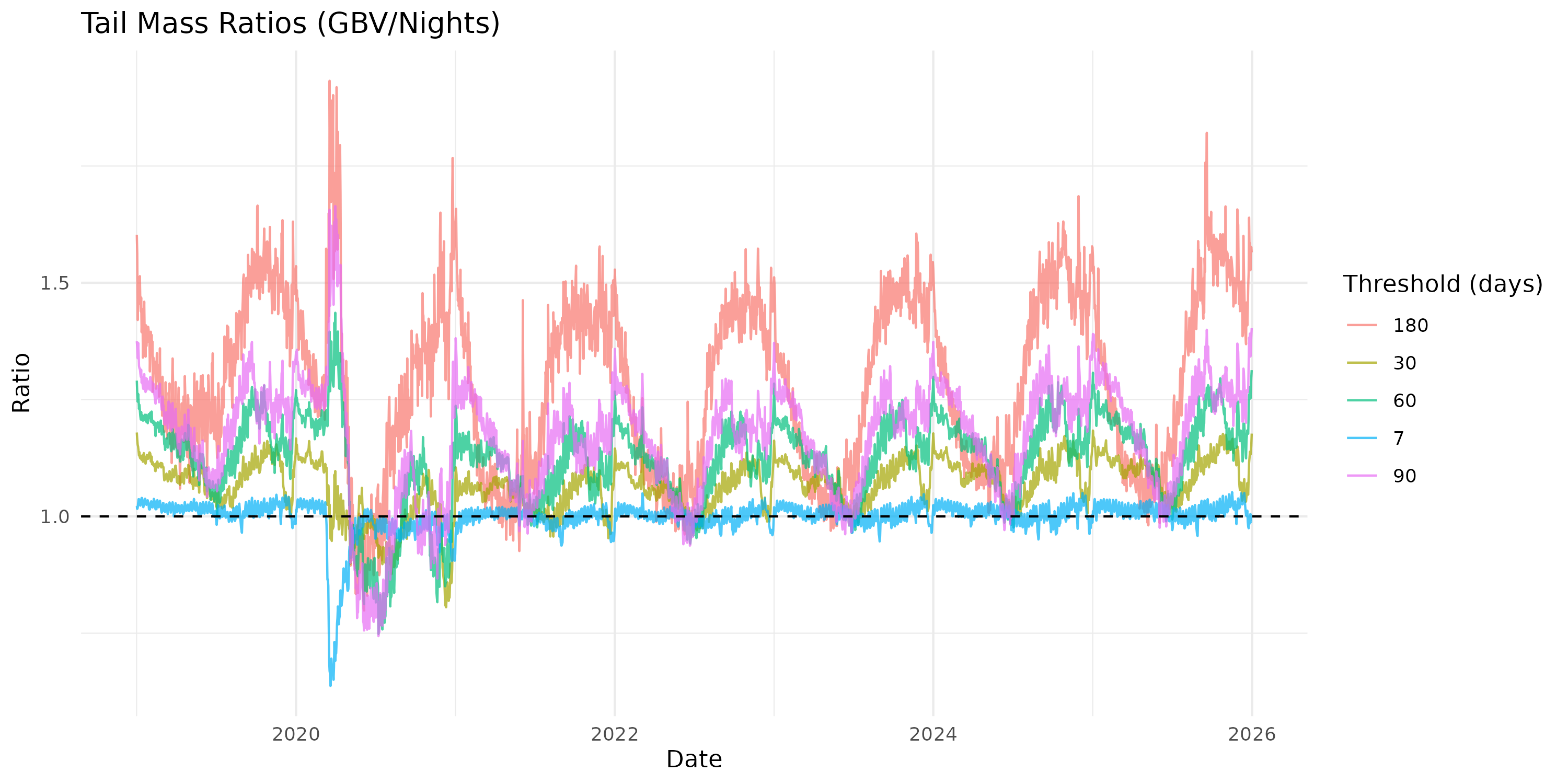}
    \caption{Tail-mass ratios (GBV/Nights) at thresholds 7, 30, 60, 90, and 180 days. At 7 days, the ratio hovers near 1.0. At longer horizons, GBV increasingly dominates: ratios reach 1.2--1.4 at 90 days and 1.5+ at 180 days during summer peaks. COVID (early 2020) briefly inverts the pattern.}
    \label{fig:tail_ratios}
\end{figure}

Figure~\ref{fig:tail_ratios} presents time series of tail-mass ratios for thresholds at 7, 30, 60, 90, and 180 days. The patterns differ markedly by threshold:

\paragraph{7 days.} The ratio hovers near 1.0 with modest variation. There is no systematic dominance of GBV over Nights at very short horizons. This makes sense: same-week bookings are often spontaneous or necessity-driven, with less differentiation between high-value and average travelers.

\paragraph{30 days.} GBV begins to pull ahead. The ratio typically ranges from 1.05 to 1.15, with seasonal peaks reaching 1.20. The gap is small but consistent.

\paragraph{60 days.} The ratio increases to 1.10--1.25, with summer peaks near 1.30. The mid-range horizon is where the volume-revenue divergence concentrates.

\paragraph{90 days.} Ratios of 1.20--1.40 are common, with peaks exceeding 1.50 in some summers. Beyond three months, GBV's tail-mass advantage becomes substantial.

\paragraph{180 days.} The ratio is most volatile here, ranging from 1.0 to 1.75. Summer peaks are pronounced; winter troughs can approach 1.0. The six-month-plus horizon captures relatively few bookings, so daily ratios are noisier.

COVID-19 disrupted these patterns. In early 2020, as booking windows collapsed, the ratios at longer thresholds dropped sharply and briefly inverted (GBV $<$ Nights) for some periods. Recovery restored the pre-pandemic pattern, though with higher baseline ratios in the post-2022 period.

\subsubsection{Absolute Tail Mass}

\begin{figure}[ht!]
    \centering
    \includegraphics[width=0.95\textwidth]{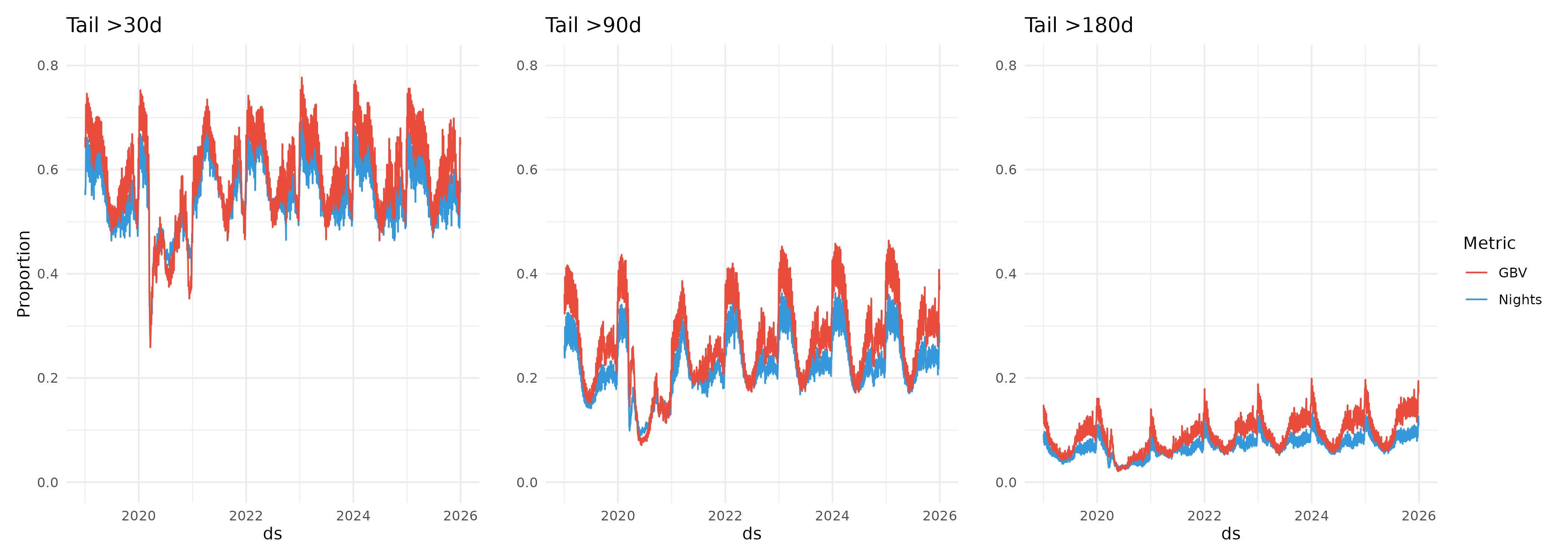}
    \caption{Proportion of daily bookings beyond 30, 90, and 180 days for Nights (blue) and GBV (red). GBV exceeds Nights at all thresholds. COVID shows a collapse in tail mass for both metrics as booking windows shortened dramatically.}
    \label{fig:tail_mass_comparison}
\end{figure}

Figure~\ref{fig:tail_mass_comparison} shows the absolute proportion of daily bookings beyond 30, 90, and 180 days for each metric.

At the 30-day threshold, roughly 55--70\% of bookings (by either measure) fall beyond this horizon, depending on season. GBV exceeds Nights by 5--10 percentage points consistently. At the 90-day threshold, 20--40\% of bookings remain in the tail, with GBV again exceeding Nights. At 180 days, only 5--15\% of bookings are this far out, but GBV maintains its relative advantage.

The COVID period shows a dramatic collapse in tail mass for both metrics: in spring 2020, the share of bookings beyond 90 days fell from 30\% to under 10\% as travelers stopped planning ahead. Recovery was gradual, with tail mass returning to pre-pandemic levels by mid-2022.

These patterns confirm that the volume-revenue divergence is real and not merely an artifact of ratio construction. GBV genuinely concentrates more heavily in mid-range horizons than Nights, and this concentration persists across seasons and regimes.

\subsection{GPD Tail Estimates}

\subsubsection{Shape Estimates by Threshold}

\begin{figure}[ht!]
    \centering
    \includegraphics[width=0.65\textwidth]{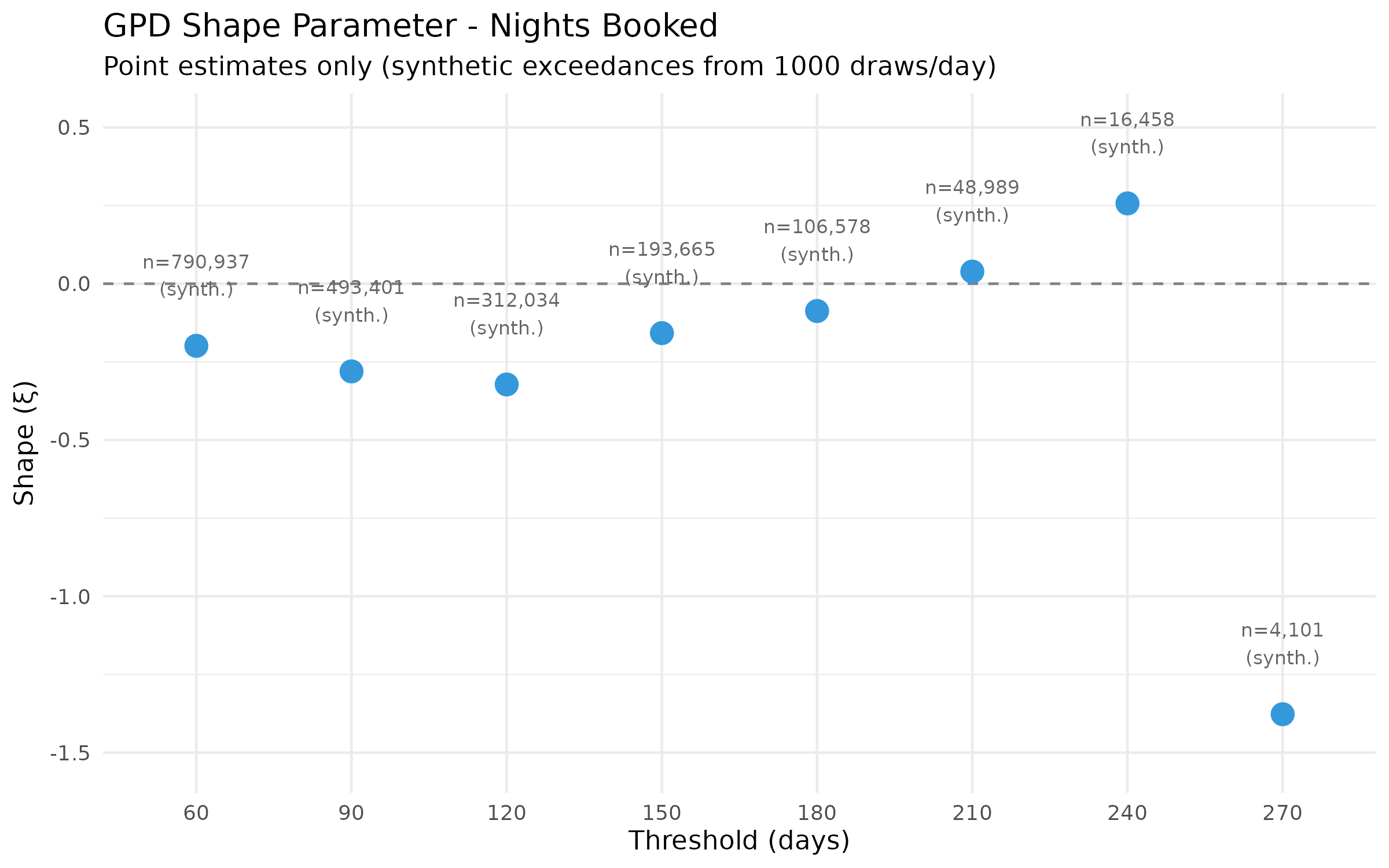}
    \caption{GPD shape parameter $\xi$ for Nights at thresholds 60--270 days. Counts shown are synthetic exceedances (from 1,000 draws/day pooled across all days), not raw booking counts. Negative values suggest bounded tails within the observed window. Estimates are stable for $u \leq 150$ days; above, truncation effects dominate.}
    \label{fig:gpd_nights_shape}
\end{figure}

\begin{figure}[ht!]
    \centering
    \includegraphics[width=0.65\textwidth]{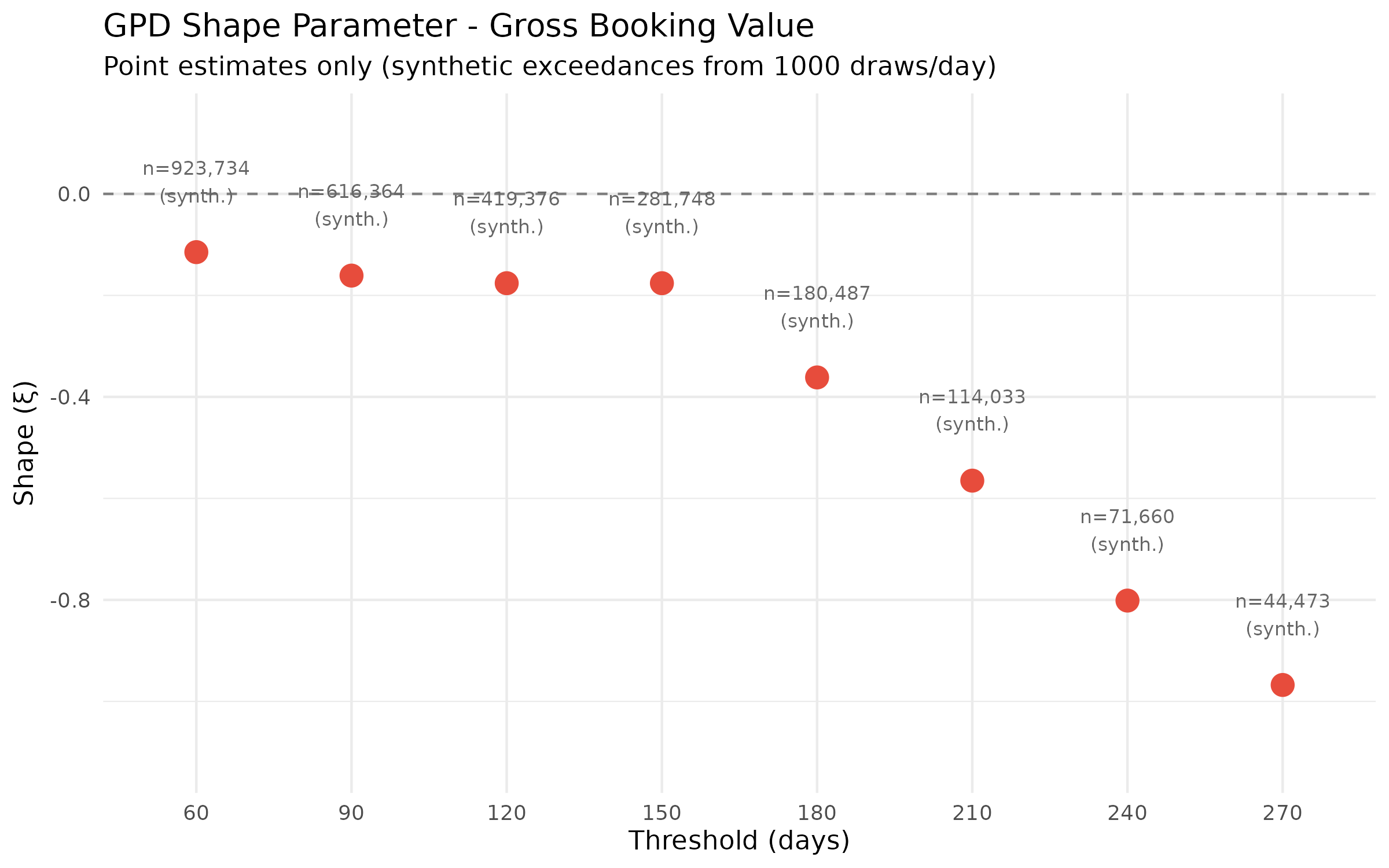}
    \caption{GPD shape parameter $\xi$ for GBV at thresholds 60--270 days. Counts are synthetic exceedances. GBV shows less negative $\xi$ than Nights at thresholds 60--120 days ($\approx -0.15$ vs.\ $-0.25$), suggesting marginally lighter tail decay; the gap narrows by 150 days. High-threshold instability is pronounced.}
    \label{fig:gpd_gbv_shape}
\end{figure}

Figures~\ref{fig:gpd_nights_shape} and \ref{fig:gpd_gbv_shape} present GPD shape parameter estimates $\hat{\xi}$ at thresholds from 60 to 270 days, with exceedance counts.

For Nights, shape estimates at low thresholds are negative and relatively stable: $\hat{\xi} \approx -0.20$ at $u = 60$, $-0.28$ at $u = 90$, $-0.32$ at $u = 120$. These values indicate bounded (Weibull-type) tails, consistent with the practical constraint that travelers rarely book more than a year in advance. At $u = 150$, the estimate is $-0.16$, still negative but less extreme.

For GBV, low-threshold estimates are also negative but less so: $\hat{\xi} \approx -0.12$ at $u = 60$, $-0.16$ at $u = 90$, $-0.18$ at $u = 120$ and $u = 150$. The less negative values suggest that GBV's tail decays more slowly than Nights' in the mid-range, consistent with the tail-mass ratio findings.

\subsubsection{Threshold Stability}

\begin{figure}[ht!]
    \centering
    \includegraphics[width=0.75\textwidth]{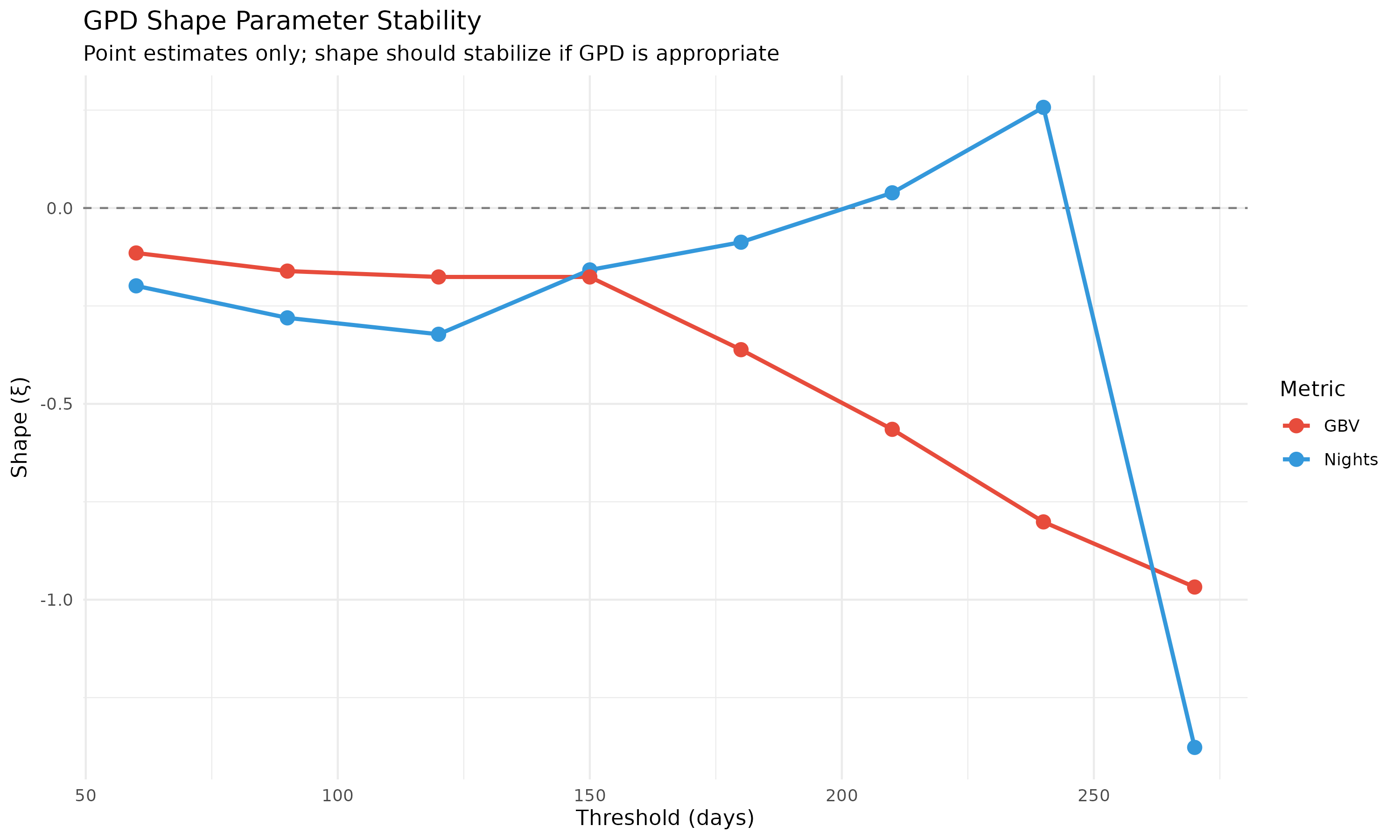}
    \caption{GPD shape parameter stability across thresholds. Dashed line marks $\xi = 0$. Below 150 days, both metrics show stable $\xi \in (-0.3, -0.1)$, consistent with bounded tails. Above 150 days, estimates diverge: Nights swings positive then collapses to $-1.4$; GBV decreases monotonically to $-1.0$. This instability reflects right-truncation at 365 days.}
    \label{fig:gpd_stability}
\end{figure}

Figure~\ref{fig:gpd_stability} plots $\hat{\xi}$ against threshold for both metrics. The pattern is striking:

\paragraph{Below 150 days.} Both metrics show relatively stable estimates in the range $\xi \in (-0.3, -0.1)$. GBV is generally less negative than Nights at thresholds 60--120 days, implying lighter tail decay; the difference narrows by 150 days. The stability suggests that GPD is a reasonable model for this range.

\paragraph{Above 150 days.} Estimates diverge dramatically. For Nights, $\hat{\xi}$ increases through zero around $u = 200$, peaks at $+0.26$ at $u = 240$ (implying heavy tails, clearly implausible), then crashes to $-1.38$ at $u = 270$. For GBV, $\hat{\xi}$ decreases monotonically, reaching $-0.97$ at $u = 270$.

This instability is a truncation artifact. The data are right-truncated at 365 days: we observe no lead times beyond this boundary. As the threshold approaches 365, the GPD estimator ``sees'' an artificial upper bound and interprets it as extreme tail boundedness ($\xi \ll 0$). But the path to this boundary is erratic because the estimator is extrapolating from a shrinking and distorted sample of exceedances.

\paragraph{Recommendation.} We restrict GPD inference to thresholds $\leq 150$ days, where estimates are stable and sample sizes adequate. Table~\ref{tab:gpd_results} reports these estimates. Within this range, both metrics show negative shape parameters, with GBV exhibiting less negative values at 60--120 days ($\xi \approx -0.15$ vs.\ $-0.25$ for Nights); the gap narrows by 150 days. This pattern is consistent with GBV having lighter tail decay in the mid-range. However, because the support is bounded at 365 by construction, negative $\xi$ does not necessarily imply a structural upper bound on booking behavior; it may simply reflect the truncation.

\subsection{Parametric Distribution Fits}

\subsubsection{Parameter Estimates}

Table~\ref{tab:param_summary} reports average parameter estimates across all days for the three parametric families.

For Gamma, the shape parameter averages 0.77 for Nights and 0.80 for GBV. Values below 1 indicate right-skewed distributions with mode at zero, consistent with the concentration of bookings at short lead times. The rate parameter averages 0.013 for Nights and 0.012 for GBV. The combination of higher shape and lower rate for GBV implies a distribution shifted rightward, toward longer lead times.

For Weibull, shape averages 0.85 for Nights and 0.87 for GBV; scale averages 54.2 and 61.7 respectively. The pattern is similar: GBV's parameters imply a distribution with more mass at longer horizons.

For Lognormal, $\mu$ averages 3.41 for Nights and 3.52 for GBV; $\sigma$ averages 1.32 for both. The higher $\mu$ for GBV again indicates a rightward shift.

Across all three families, the parameter differences between Nights and GBV point in the same direction: revenue-weighted bookings are distributed further into the future than volume-weighted bookings. This consistency across distributional assumptions strengthens the substantive finding.

\subsubsection{Best-Fit Distribution by Cross-Entropy}

Table~\ref{tab:wins_summary} tallies which distribution achieves the lowest cross-entropy on each day.

For Nights: Gamma wins 61.4\% of days (1,570 of 2,557), Weibull wins 37.9\% (969), and Lognormal wins 0.7\% (18). For GBV: Gamma wins 51.9\% (1,327), Weibull wins 45.4\% (1,160), and Lognormal wins 2.7\% (70).

The competition between Gamma and Weibull is close, especially for GBV. Lognormal rarely wins. This pattern holds across subperiods: Gamma's edge is consistent in pre-COVID, COVID, and post-COVID regimes.

\subsubsection{Cross-Entropy Differences}

\begin{figure}[ht!]
    \centering
    \includegraphics[width=0.95\textwidth]{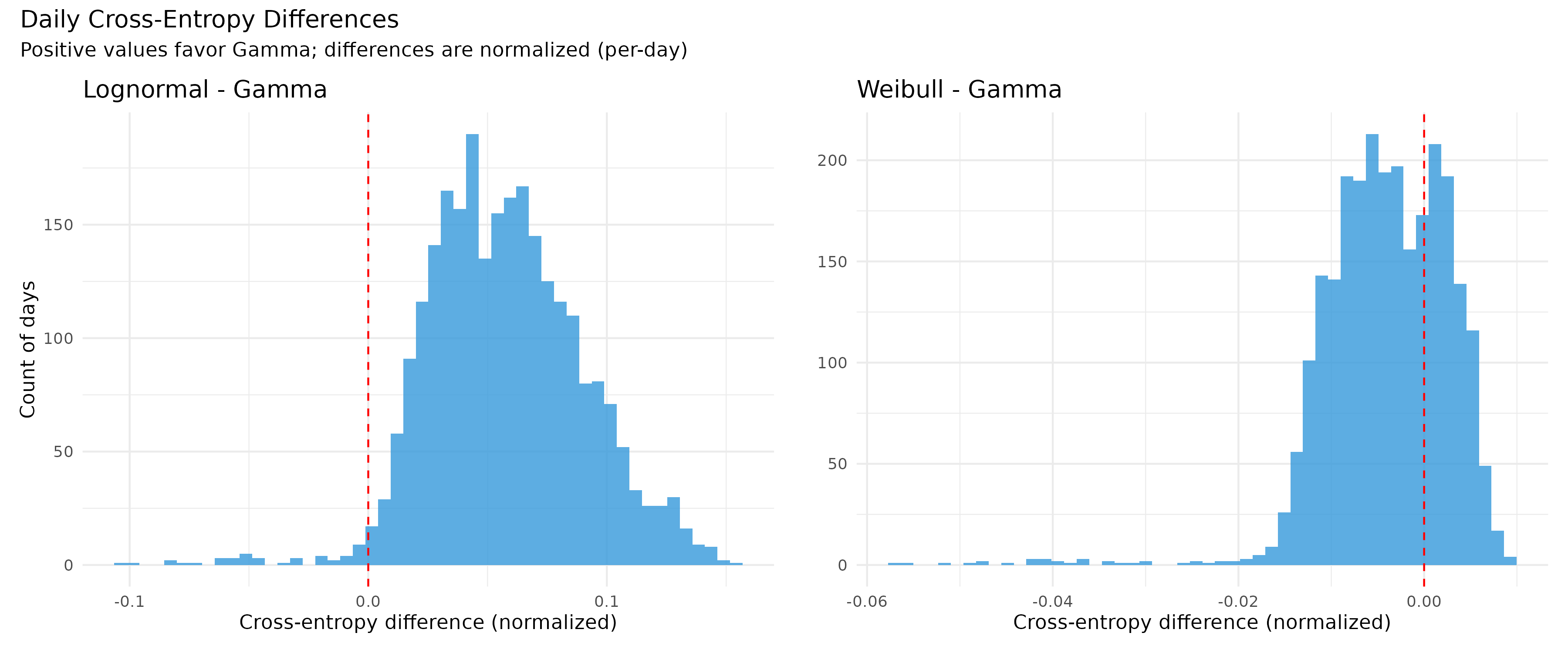}
    \caption{Histograms of daily cross-entropy differences: Lognormal minus Gamma (left), Weibull minus Gamma (right). Positive values favor Gamma. These are normalized (per-day) differences; they would scale with booking counts if counts were available. The Lognormal--Gamma difference is almost entirely positive (0.02--0.20), confirming Gamma's dominance. The Weibull--Gamma difference is tightly centered around zero ($\pm 0.02$): these distributions are effectively interchangeable.}
    \label{fig:threeway_aic}
\end{figure}

Figure~\ref{fig:threeway_aic} shows histograms of daily cross-entropy differences. These are normalized (per-day) values; they would scale with booking counts if counts were available.

The Lognormal--Gamma difference is almost entirely positive, ranging from 0.02 to 0.20. This confirms Gamma's decisive advantage over Lognormal: on virtually every day, Gamma fits better.

The Weibull--Gamma difference is tightly centered around zero, with most values in $(-0.02, +0.02)$. The distributions are effectively interchangeable for this application. Gamma's slight edge in win percentage reflects many days where the difference is tiny and favors Gamma by chance.

The practical implication is that either Gamma or Weibull provides an adequate two-parameter summary of daily lead-time pmfs. Lognormal's heavier tail does not match the bounded nature of the data.

\subsection{Parametric vs.\ Nonparametric}

Table~\ref{tab:scores_summary} compares Gamma and GAM fits using CRPS and KLD.

GAM achieves substantially lower CRPS: 0.0007 vs.\ 0.056 for Nights (an 80$\times$ improvement) and 0.008 vs.\ 0.157 for GBV (an 18$\times$ improvement). These extreme ratios reflect the GAM's flexibility: with 10--20 effective degrees of freedom, it approximates a smoothed empirical pmf rather than imposing a two-parameter shape. The GAM captures fine-grained features (local bumps, asymmetric tails, multimodal tendencies on some days) that the Gamma cannot match. These are in-sample scores: each GAM is fit and evaluated on the same day's pmf. The GAM CRPS should be interpreted as a lower bound on achievable in-sample fit with flexible smoothers, not as a forecast performance claim. Out-of-sample evaluation would be needed to assess whether GAM's flexibility translates to better forecasts or merely captures noise.

KLD shows a similar but less dramatic pattern: GAM achieves 0.016 vs.\ Gamma's 0.039 for Nights, and 0.121 vs.\ 0.157 for GBV.

This creates a parsimony-flexibility tradeoff. Gamma wins by cross-entropy when parameter counts are equal (all families have $k=2$). GAM wins by CRPS and KLD because it has more effective parameters. For forecasting applications requiring well-calibrated predictive distributions, GAM may be preferred, but this claim requires out-of-sample validation. For descriptive purposes where interpretability matters (``what is the typical shape parameter?''), Gamma provides a compact summary.

\section{Discussion}
\label{sec:discussion}

\subsection{Main Findings}

The central finding is that revenue-weighted bookings concentrate more heavily in the 30--90 day horizon than volume-weighted bookings. The tail-mass ratios quantify this: beyond 90 days, GBV typically exceeds Nights by 20--50\%, with ratios reaching 75\% at the 180-day threshold during peak seasons. This is not an extreme-tail phenomenon. Both metrics have bounded tails beyond 300 days; this is a mid-range shift. Guests who generate higher revenue tend to book earlier, but not dramatically earlier.

The structural breaks in Wasserstein distance show that this gap is not constant. COVID compressed it; recovery restored it; seasonal patterns modulate it year after year. The five identified breakpoints partition the sample into regimes with distinct divergence levels, confirming that the volume-revenue relationship evolves with external conditions.

The distributional fitting comparison yields a practical conclusion: Gamma and Weibull are both adequate, with Gamma holding a slight edge. Lognormal is dominated. GAMs fit better by CRPS but sacrifice interpretability; their in-sample advantage reflects flexibility, not necessarily forecast skill. For most applications, a Gamma distribution with shape $\approx 0.8$ and rate $\approx 0.013$ provides a reasonable two-parameter summary of Airbnb lead-time distributions.

The GPD analysis offers a methodological caution. Shape estimates are relatively stable below 150 days, where both metrics show $\xi \approx -0.15$ to $-0.25$. Within the observed window, this suggests bounded tails, but because the data are truncated at 365 days by construction, a negative shape parameter is largely tautological. We interpret the bounded-tail signal as descriptive of the observed distribution rather than a structural claim about booking behavior beyond the platform's booking window. Above 150 days, instability induced by truncation would mislead inference if ignored. Researchers applying GPD to truncated data should perform threshold stability diagnostics before reporting point estimates.

\subsection{Implications for Tourism Practitioners}

The volume-revenue divergence has concrete consequences for several areas of tourism and hospitality management.

\paragraph{Revenue forecasting and cash-flow planning.} Models that treat Nights and GBV as interchangeable, or that forecast GBV by scaling a Nights forecast, will systematically mispredict revenue timing. Specifically, they will underestimate revenue from mid-range bookings (one to three months ahead) and overestimate revenue from short-horizon bookings (within one month).

To illustrate the magnitude: suppose a forecasting model predicts that 30\% of next month's Nights will be booked more than 60 days in advance. If the model assumes GBV follows the same distribution, it will predict 30\% of revenue in this category. But if the true GBV tail mass is 36\% (a ratio of 1.2, typical of non-summer months), the model will underpredict far-horizon revenue by 6 percentage points. During summer peaks, when the ratio rises to 1.3 or higher, the bias reaches 9 percentage points. Over a fiscal quarter, these errors compound as each month's forecast feeds into cumulative cash-flow projections, producing systematic under-reporting of revenue already ``in the pipeline'' from advance bookings and over-reliance on short-horizon bookings that are inherently less certain.

\paragraph{Dynamic pricing.} Pricing algorithms that use volume-based booking curves to estimate remaining demand may underweight the 30--90 day window where revenue disproportionately concentrates. If a pricing tool observes that ``only 25\% of room nights are booked 60+ days out'' and treats this as a signal of weak future demand, it may set prices too low, failing to recognize that a higher share of \textit{revenue} is already committed at that horizon. Adjusting the demand signal by the tail-mass ratio could improve price-setting in the mid-range without requiring a full model overhaul.

\paragraph{Cancellation policy and risk management.} Cancellation exposure is inherently tied to lead time: bookings made further in advance face a longer window during which travelers may cancel. Because high-value bookings skew toward longer lead times, the revenue at risk from cancellations may be larger than a volume-based analysis would suggest. Revenue managers assessing cancellation reserves or evaluating the cost of flexible cancellation policies should account for the fact that the lead-time distribution of revenue differs from that of volume. In practice, this means that the ``average'' cancellation window, weighted by revenue, is longer than the average cancellation window weighted by booking count.

\paragraph{Post-shock recalibration.} The structural breaks we identify imply that the volume-revenue relationship is not a fixed feature of the market. COVID-19 compressed the divergence; recovery amplified it above pre-pandemic levels. Models trained on pre-2020 data will embed a systematically different volume-revenue relationship than what prevails in 2023--2025. Tourism operators and platform designers should periodically reassess the volume-revenue divergence, especially after major market disruptions, rather than treating it as a stable calibration parameter.

\paragraph{A simple correction.} For practitioners seeking a quick adjustment rather than a full model rebuild, the tail-mass ratio provides a straightforward correction factor. Multiplying the volume-based forecast share for the 30--90 day window by the current tail-mass ratio (approximately 1.2--1.4 at the 90-day threshold) reallocates forecasted revenue toward mid-range horizons. At current ratios, this adjustment shifts 2--4 percentage points of forecasted revenue from short-horizon to mid-range bins, reducing cash-flow timing bias with minimal model complexity.

\paragraph{Destination management and tourism planning.} The volume-revenue divergence also has implications beyond the firm level. Destination marketing organizations (DMOs) and tourism boards that rely on arrival counts or room-night statistics to forecast economic impact may underestimate the contribution of mid-range advance bookings, which carry disproportionate revenue. If a region's tourism strategy emphasizes last-minute promotional campaigns to fill occupancy gaps, the revenue uplift from those bookings will be smaller per unit than what the region already has committed at the 30--90 day horizon. Conversely, policies or events that attract early-planning, high-spending travelers (such as festivals announced well in advance or convention bookings) may generate a larger revenue contribution than their share of room nights would suggest. The distributional tools used here could help DMOs move beyond headcount-based indicators toward revenue-aware assessments of tourism demand.

\paragraph{Generalizability to the broader accommodation sector.} Although our data come from a single platform, the underlying mechanism (higher-spending guests booking earlier) is not specific to Airbnb. Hotels with diverse room categories, resorts with premium packages, and serviced-apartment platforms all feature traveler segments with different willingness to pay and different planning horizons. The volume-revenue divergence documented here is likely a general feature of accommodation markets, with the magnitude varying by market segment, destination type, and platform structure. Replication across accommodation types would clarify where the divergence is largest and where volume-based proxies remain adequate.

\subsection{Implications for Tail Inference}

The GPD instability we document is a general problem for truncated data, not specific to lead times. Any dataset with a hard upper bound will exhibit similar artifacts when GPD is applied near that bound. The threshold stability diagnostic is essential: if $\hat{\xi}$ does not stabilize, the model is unreliable in that range.

Our recommendation, restricting inference to $u \leq 150$ days when the truncation is at 365, is conservative. Other analysts might choose a higher cutoff if their data extend further or their sample sizes are larger. The key is to check stability empirically rather than relying on asymptotic guarantees. This finding is relevant beyond hospitality: any bounded duration variable (e.g., subscription lengths capped by contract terms, event lead times bounded by announcement dates) will face analogous truncation artifacts in extreme-value analysis.

\subsection{Limitations}

Several limitations warrant mention.

First, the analysis is in-sample. We fit distributions to observed data but do not evaluate out-of-sample forecast accuracy. The GAM's superior CRPS might not translate to better forecasts if the model overfits daily idiosyncrasies.

Second, we analyze a single region. Lead-time patterns likely differ across geographies: European travelers may book further ahead than North Americans; urban destinations may have shorter booking windows than leisure destinations. Generalizing our findings requires replication in other markets.

Third, we omit covariates. External factors (macroeconomic conditions, competitor pricing, local events) surely influence lead-time distributions. Incorporating these in a regression framework could explain some of the day-to-day variation we observe.

Fourth, the 365-day truncation, while capturing 99.5\% of bookings, precludes analysis of the true extreme tail. Travelers who book 18+ months ahead are rare but may be substantively different; we cannot characterize them with this data.

Fifth, although we demonstrate that the volume-revenue divergence is substantial and systematic, we do not identify the causal mechanism. The divergence could arise because higher-spending guests inherently plan further ahead, because premium properties are listed earlier, because price discrimination over the booking curve concentrates revenue at longer horizons, or some combination of these. Disentangling these channels would require guest-level or listing-level data, which we do not analyze here.

\subsection{Future Directions}

Several extensions are natural.

Hierarchical models could pool information across regions, improving estimates for small markets while allowing region-specific deviations. A Bayesian framework with shrinkage priors would be suitable \citep{KATZ20241556}.

Covariates could be incorporated via regression on the distribution parameters. For instance, Gamma shape and rate could depend on day-of-week, season, or economic indicators. This would connect the distributional analysis to causal questions about what drives lead-time behavior.

Out-of-sample evaluation would test whether GAM's in-sample CRPS advantage translates to forecast improvement. A rolling-origin design with multiple forecast horizons would be informative.

Guest-level or listing-level analysis could illuminate the mechanisms behind the volume-revenue divergence, testing whether the pattern is driven by traveler heterogeneity (high-spending guests booking earlier), property heterogeneity (premium listings attracting earlier bookings), or price-discrimination effects (prices increasing closer to check-in). Such analysis would also clarify whether the divergence is specific to Airbnb or generalizes to hotels and other accommodation types.

Finally, extension to other compositional outcomes (not just lead times but geographic mix, property-type mix, or price-tier mix) would broaden the applicability of these methods to other questions in tourism analytics.

\section{Conclusion}

We have analyzed daily lead-time distributions for Nights Booked and Gross Booking Value on Airbnb, treating each day's allocation as a compositional vector. The analysis yields three main findings.

First, GBV concentrates more heavily in mid-range horizons than Nights. Beyond 90 days, GBV typically exceeds Nights by 20--50\%, with ratios reaching 75\% at the 180-day threshold during peak seasons. This is a mid-range shift, not an extreme-tail phenomenon.

Second, Gamma and Weibull fit daily pmfs comparably well, with Gamma winning 55--60\% of days by cross-entropy. Lognormal is dominated. GAMs achieve superior in-sample CRPS but sacrifice interpretability.

Third, GPD confirms bounded tails ($\xi < 0$) at reliable thresholds ($\leq 150$ days), but threshold instability at higher values reflects truncation artifacts. Researchers should perform stability diagnostics before reporting GPD estimates on truncated data.

These findings contribute to tourism forecasting methodology and practical revenue management. The documented volume-revenue divergence, and its evolution over COVID and recovery, has direct implications for cash-flow prediction, dynamic pricing, and cancellation risk assessment. For tourism researchers, the key empirical contribution is that volume-based demand indicators and revenue-based demand indicators carry systematically different temporal information: studies that treat room nights as a proxy for revenue, or vice versa, risk mischaracterizing the economic timing of tourist demand. For practitioners, the key takeaway is that a simple tail-mass ratio correction can reduce cash-flow forecasting bias by reallocating predicted revenue from short-horizon to mid-range bins, requiring no model overhaul. The methodological lessons (interval-censored likelihood, CRPS scoring, threshold stability checks) apply broadly to compositional and duration data in tourism and beyond, and the distributional comparison framework demonstrated here could be extended to other dimensions of tourism demand, including geographic mix, traveler-segment composition, and length-of-stay distributions.



\clearpage

\bibliographystyle{chicago}
\bibliography{references}

\clearpage

\section*{Figures}

\clearpage

\section*{Tables}

\begin{table}[ht!]
\centering
\caption{Average parameter estimates across all 2,557 days for three parametric families. GBV shows higher Gamma shape, lower rate, higher Weibull scale, and higher Lognormal $\mu$ than Nights, all indicating a distribution shifted toward longer lead times.}
\label{tab:param_summary}
\begin{tabular}{lcccccc}
\toprule
\textbf{Metric} & \textbf{LN $\mu$} & \textbf{LN $\sigma$} & \textbf{Wei shape} & \textbf{Wei scale} & \textbf{Gamma shape} & \textbf{Gamma rate}\\
\midrule
GBV    & 3.52 & 1.32 & 0.87 & 61.7 & 0.80 & 0.012 \\
Nights & 3.41 & 1.32 & 0.85 & 54.2 & 0.77 & 0.013 \\
\bottomrule
\end{tabular}
\end{table}

\begin{table}[ht!]
\centering
\caption{Distribution of best-fitting models by cross-entropy. Gamma wins most often, especially for Nights. Weibull is close behind for GBV. Lognormal rarely wins.}
\label{tab:wins_summary}
\begin{tabular}{llcc}
\toprule
\textbf{Metric} & \textbf{Best Fit} & \textbf{Days} & \textbf{\%}\\
\midrule
Nights & Gamma     & 1,570 & 61.4 \\
Nights & Weibull   & 969   & 37.9 \\
Nights & Lognormal & 18    & 0.7 \\
\midrule
GBV    & Gamma     & 1,327 & 51.9 \\
GBV    & Weibull   & 1,160 & 45.4 \\
GBV    & Lognormal & 70    & 2.7 \\
\bottomrule
\end{tabular}
\end{table}

\begin{table}[ht!]
\centering
\caption{CRPS and KLD for Gamma vs.\ GAM. Lower is better. GAM achieves 18--80$\times$ lower CRPS than Gamma, reflecting its flexibility, but uses 10--20 effective degrees of freedom vs.\ Gamma's 2 parameters.}
\label{tab:scores_summary}
\begin{tabular}{lcccc}
\toprule
 & \textbf{Nights (Gamma)} & \textbf{Nights (GAM)} & \textbf{GBV (Gamma)} & \textbf{GBV (GAM)}\\
\midrule
CRPS & 0.056 & 0.0007 & 0.157 & 0.008 \\
KLD  & 0.039 & 0.016  & 0.157 & 0.121 \\
\bottomrule
\end{tabular}
\end{table}

\begin{table}[ht!]
\centering
\caption{GPD shape parameter estimates at reliable thresholds ($u \leq 150$ days). Negative $\xi$ suggests bounded tails within the observed window, though this may partly reflect truncation at 365 days. GBV shows less negative $\xi$ than Nights at 60--120 days; the difference narrows by 150. Counts $n$ are synthetic exceedances (1,000 draws/day $\times$ 2,557 days, pooled); we report point estimates but not confidence intervals (see text).}
\label{tab:gpd_results}
\begin{tabular}{lcccc}
\toprule
\textbf{Threshold} & \textbf{Nights $\xi$} & \textbf{Nights $n$ (synth.)} & \textbf{GBV $\xi$} & \textbf{GBV $n$ (synth.)}\\
\midrule
60 days  & $-0.20$ & 790,937 & $-0.12$ & 923,734 \\
90 days  & $-0.28$ & 493,401 & $-0.16$ & 616,364 \\
120 days & $-0.32$ & 312,034 & $-0.18$ & 419,376 \\
150 days & $-0.16$ & 193,665 & $-0.18$ & 281,748 \\
\bottomrule
\end{tabular}
\end{table}

\end{document}